\documentclass[]{raa}          

\usepackage{graphicx,times}
\usepackage{natbib}
\usepackage{amssymb,amsmath}
\usepackage[flushleft]{threeparttable}
\bibpunct{(}{)}{;}{a}{}{,}

\usepackage[pagebackref=true]{hyperref}
\newcommand{\kms}{km s$^{-1}$}
\begin{document}

   \title{A baseline correction algorithm for FAST}

 \volnopage{ {\bf 20XX} Vol.\ {\bf X} No. {\bf XX}, 000--000}
   \setcounter{page}{1}

   \author{De-Jian Liu
   \inst{1, 2}, Ye Xu\inst{1}, Ying-Jie Li\inst{1}, Ze-Hao Lin
      \inst{1, 2},  Shuai-Bo Bian\inst{1, 2}, Chao-Jie Hao \inst{1, 2}
   }

   \institute{Purple Mountain Observatory, Chinese Academy of Sciences, Nanjing 210008, China; {\it xuye@pmo.ac.cn}\\
        \and
             School of Astronomy and Space Science, University of Science and Technology of China, Hefei 230026, PR China\\
\vs \no
   {\small Received 20XX Month Day; accepted 20XX Month Day}
}

\abstract{The Five-hundred-meter Aperture Spherical radio Telescope (FAST) is the most sensitive ground-based, single-dish radio telescope on Earth. However, the original H~\textsc{i} spectra produced by FAST are affected by standing waves. To maximize the power of FAST for high-sensitivity observations, we proposed an algorithm that combines fast Fourier transforms and extreme envelope curves to automatically correct the baselines of FAST H~\textsc{i} spectra and remove standing waves from the baselines. This algorithm can reduce the amplified noise level caused by standing waves to a near-ideal level without losing signals or introducing false signals. The root mean square of the average baseline reaches $\sim$8~mK, approaching the theoretical sensitivity of an H~\textsc{i} spectrum produced by FAST for an integration time of 335 minutes, i.e., $\sim$ 6~mK.
}

   \authorrunning{D.-J. Liu et al. }            
   \titlerunning{Baseline correction of FAST}  
   \maketitle

%
\section{Introduction}
\label{sect:intro}
The Five-hundred-meter Aperture Spherical radio Telescope (FAST), located in Guizhou Province of Southwest China, is the world’s largest single-dish radio telescope. It is an important facility for surveying neutral hydrogen up to the edge of the universe, detecting weak space signals, hearing possible signals from other civilizations, etc.~\citep{Nan+etal+2011,Qian+etal+2020}. Although it has produced significant scientific achievements~\citep{Qian+etal+2020}, the baselines of the original H~\textsc{i} spectra are not flat enough and they contain massive standing waves that might be generated by reflections between the dish and the receiver cabin~\citep{Jiang+etal+2020}. Although efforts have been made to minimize the standing wave effects of FAST data~\citep{Jiang+etal+2020}, standing waves still exist in the spectra. Standing waves can amplify the noise level of an signal; e.g., ripple amplitudes of $\sim$15~mK are commonly seen~\citep{Jiang+etal+2020}, which causes the noise of the obtained high-sensitivity spectra to be higher than the theoretically predicted noise. Some studies usually require extreme high-sensitivity observations during the analysis of the spectra; e.g., searching for the stellar winds, compact high-velocity clouds, and active star-forming dwarf galaxies~\citep[][Li et al. in preparation]{Lizano+etal+1988,Giovanardi+etal+1992,Burton+etal+2001,Salzer+etal+2002}. A high-precision baseline correction method is hence necessary for studies that require extreme high-sensitivity spectra.

Polynomial fitting and trigonometric function fitting methods are commonly used to make baseline corrections~\citep{Gan+etal+2006,Baek+etal+2011}. These methods usually need to cut peaks from the original spectrum and estimate the baseline using a polynomial or trigonometric function. However, these methods can be ineffective if the baseline is complex or the format of the function is not good enough. The asymmetrically reweighted penalized least squares (arPLS) algorithm, developed from penalized least squares methods~\citep{Eilers+2003,Carlos+etal+2006,Zhang+etal+2010,Baek+etal+2015}, is a widely used baseline correction method for FAST data~\citep{Zhang+etal+2022,Wang+etal+2022}. The baseline can be estimated by changing the ``weight" parameter iteratively. Similar weights are assigned to baseline regions without peaks, while no weights or small weights are assigned for peaks; once assigned, the weights gradually reduced as the level of the signal increases. However, arPLS is not good at removing standing waves.

To correct an inclined baseline and remove standing waves from the original H~\textsc{i} spectra automatically produced by FAST, an algorithm combining fast Fourier transforms (FFTs) and extreme envelope curves (EECs), called FFTEEC, is proposed in this work.
 
\section{Data} 
\label{sect:data}
\subsection{Observations}
FAST is equipped with a 19-beam receiver and dual linear polarizer (i.e., XX and YY). The full band width of the $L$-band is 500~MHz over the frequency range 1.0--1.5~GHz. The frequency resolution of the high-resolution modes is 476.84~Hz, corresponding to a velocity resolution of $\sim 0.1$~\kms \/ @ 1.4~GHz. The beam size is 2.9$'$, and the pointing error is $\sim 0.2'$~\citep{Li+etal+2018,Jiang+etal+2019,Jiang+etal+2020}. The data used in this paper consist of H~\textsc{i} spectra of G176.51+00.20 observed on August 19th and 20th, 2021, with the 19-beam tracking observing mode, and the total integration time of the data was 335 minutes with a sampling rate of one second. We resampled the data to a velocity resolution of 0.1~\kms, and only considered data in the velocity range -7,000 to 7,000~\kms, i.e., 140\,001 channels. The data displayed in this paper are from the YY polarization of Beam M01.

\subsection{Baseline of FAST}
\label{subsect:baseline}
\begin{figure}[htbp]
    \centering
    \includegraphics[width = 7cm]{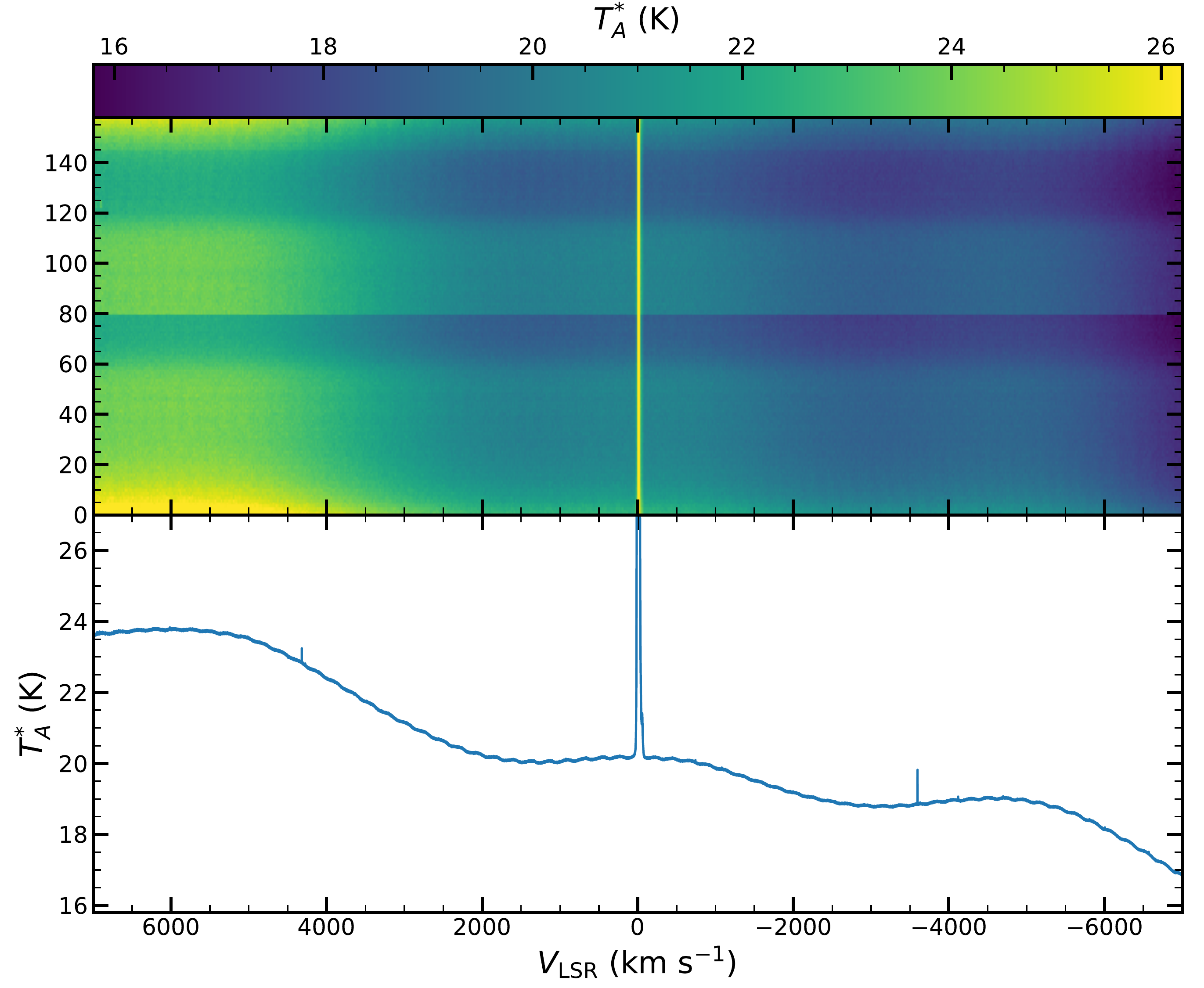}
    \includegraphics[width = 7cm]{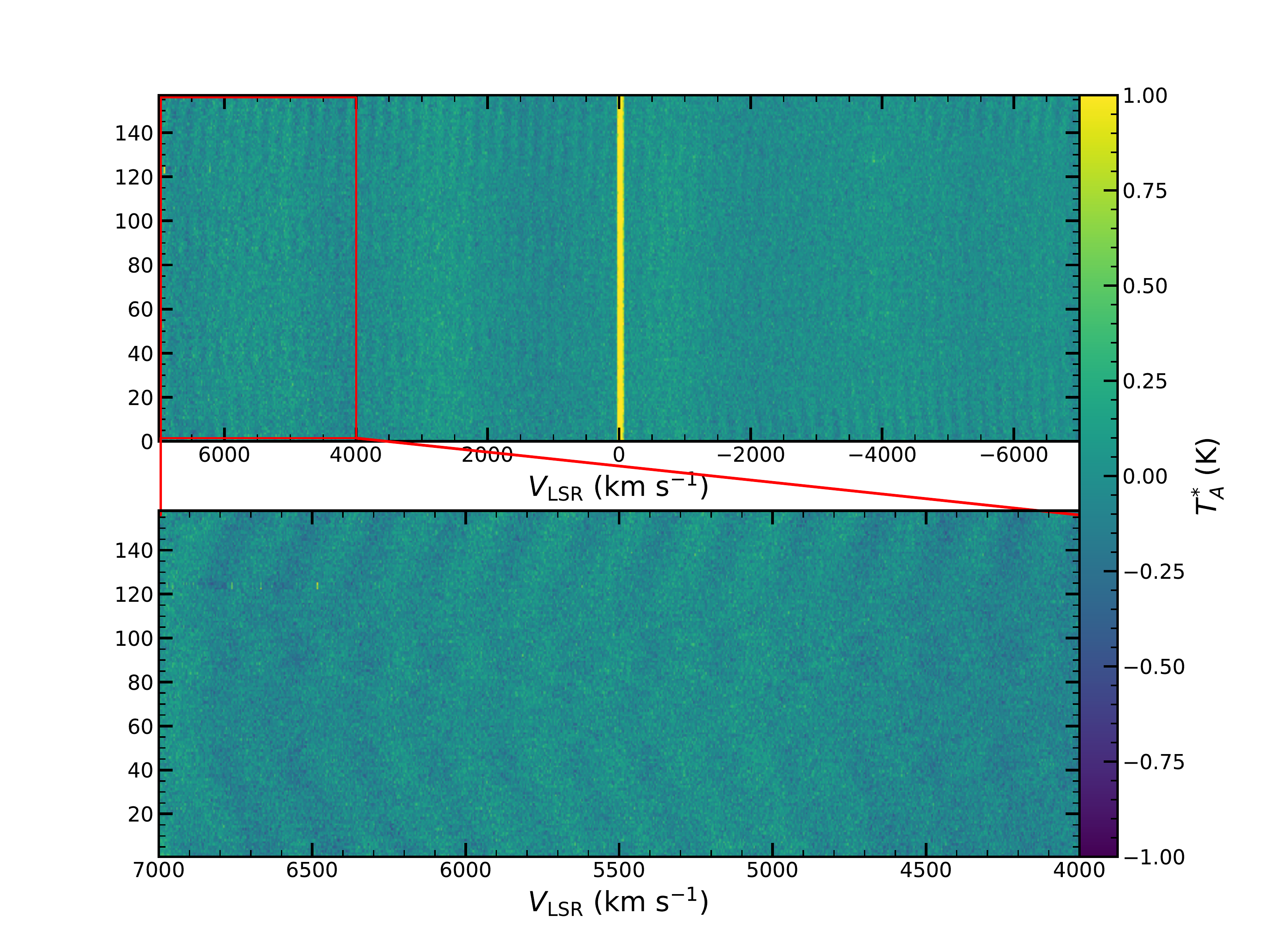}
    \caption{{\it Left}: . The water-fall image and average spectrum of original FAST spectra. The water-fall image, which contains 156 spectra, is non-uniform, which arises from the baselines of different original spectra being inconsistent. The baseline is inclined and contains massive standing waves with different frequencies. {\it Right}: Water-fall image after removing the polynomial-fit baseline and an enlarged image. The standing waves in the water-fall image are arc-shaped, indicating that they drift with time. The periods of standing waves are not stable in the different spectra.}
    \label{fig:ori}
\end{figure}

Figure~\ref{fig:ori} displays the original spectra after correcting the flux and velocity~\citep{Jiang+etal+2020}. In the left panel, the non-uniform water-fall image indicates that the baselines of the different original spectra are inconsistent, and the average spectrum with massive standing waves is inclined. The right panel shows fringes in the water-fall image after correcting the inclined baseline roughly with the polynomial fitting method. The standing waves are obvious and unstable (see the partially enlarged view). The arc-shape standing waves in the 2D water-fall image mean that the phases of the standing waves drift with time and the periods of the standing waves in different spectra are inconsistent at the same time.

\section{Method}
\label{sect:method}

\begin{figure}[htbp]
    \centering
    \includegraphics[width = 13cm]{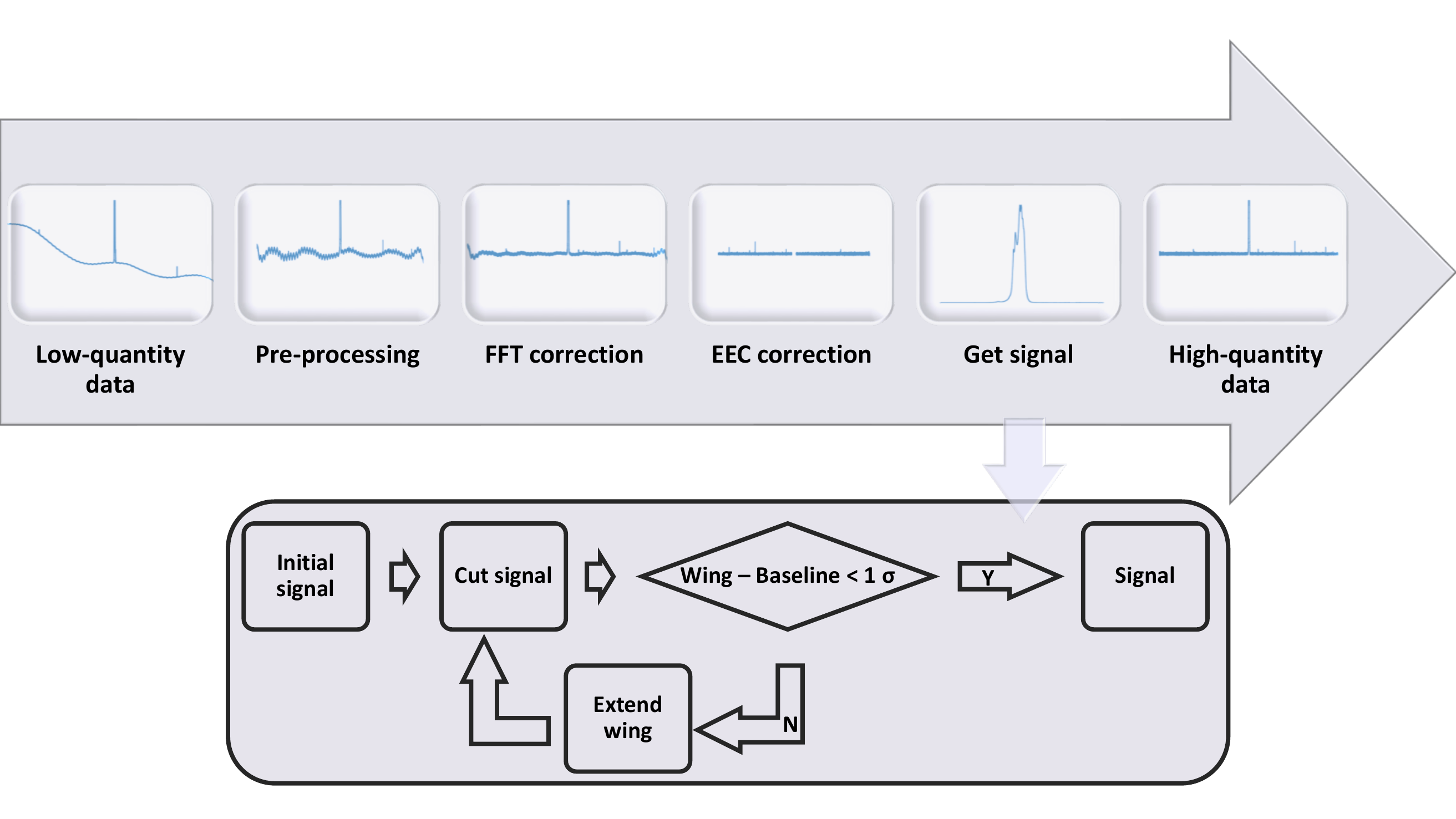}
    \caption{Pipeline of FFTEEC. The figure corresponding to each step is the average spectrum after the step is processed.}
    \label{fig:pipeline}
\end{figure}

Due to the irregular phases and periods of standing waves, as well as huge amounts of data produced by FAST, e.g., 6.0~TB for our data, we propose a highly precise baseline estimating algorithm, FFTEEC, to correct baselines automatically. First, the original spectrum is preprocessed by a polynomial fitting method. Second, the standing waves are extracted and removed using FFTs. Third, the EEC method is employed to calibrate the unsmooth parts in the baseline. Four, the extracted signal is combined with the baseline obtained in the third step as the result. Figure~\ref{fig:pipeline} displays the whole pipeline of the algorithm. 

\subsection{FFTs}
An FFT is a fast algorithm that computes discrete Fourier transforms~\citep[DFTs, ][]{Cooley+Tukey+1965}. The DFT of sequence $x$  with length $N$ can be expressed as:

\begin{equation}
    X(k) = DFT[x(n)] = \sum_{n = 1}^{N} x(n) \exp \left[ -\frac{j 2 \pi n k}{N} \right], 1 \leq k \leq N.
\end{equation}

Any frequency with a large amplitude in the frequency domain, $X$, can be considered as a standing wave. A new sequence, $Y$, is obtained by sorting $X$ from high to low and removing the first several items.  

\subsection{EECs}
An EEC, $y_{\rm e}$, can be obtained with the following steps. First, getting the smoothed sequence $y$ from $Y$, and the local maximum and minimum are extracted from $y$. Second, the maximum and minimum envelope curve of $y$, $y_{\rm max}$ and $y_{\rm min}$, respectively, can be obtained by fitting the local extrema with a linear interpolation method. Thus, $y_{\rm e}$ can be expressed as:
 
\begin{equation}
    y_{\rm e} = \frac{y_{\rm max} + y_{\rm min}}{2}.
\end{equation}

\subsection{Automatic signal extraction}
To extract a complete signal, we propose an iterative method to cut the signal automatically. The initial signal range is obtained from the EEC, where $y_{\rm e}$ is greater than 20$\sigma$ of the smoothed baseline. The signal range needs to be extended if the difference between the wing of the smoothed signal and the baseline is greater than 1$\sigma$ until it is smaller than 1$\sigma$. The method is displayed in Figure~\ref{fig:pipeline} as a subprocess that is framed by a box.

\section{Performance of FFTEEC}
\label{sect:results}
\subsection{Simulation}
\label{secsect:simulate}
The simulated data considered here contain the pure analytical signal $p$, a simulated baseline, $b$, and random noise, $j$~\footnote{The random noise was generated by the random number generator {\texttt{numpy.random.randn}} of {\texttt{python}} language.}, which can be expressed as: 

\begin{equation}
    s(n) = p(n) + b(n) + 0.1 * j(n),
\end{equation}
where $p$ is:
\begin{equation}
    p(n) = 100 \exp \left( -\frac{n^{2}}{25^{2}} \right),
\end{equation}
and $b$ contains the residuals extracted from the real data by the method in this paper.

\begin{figure}[htbp]
    \centering
    \includegraphics[width = 7cm]{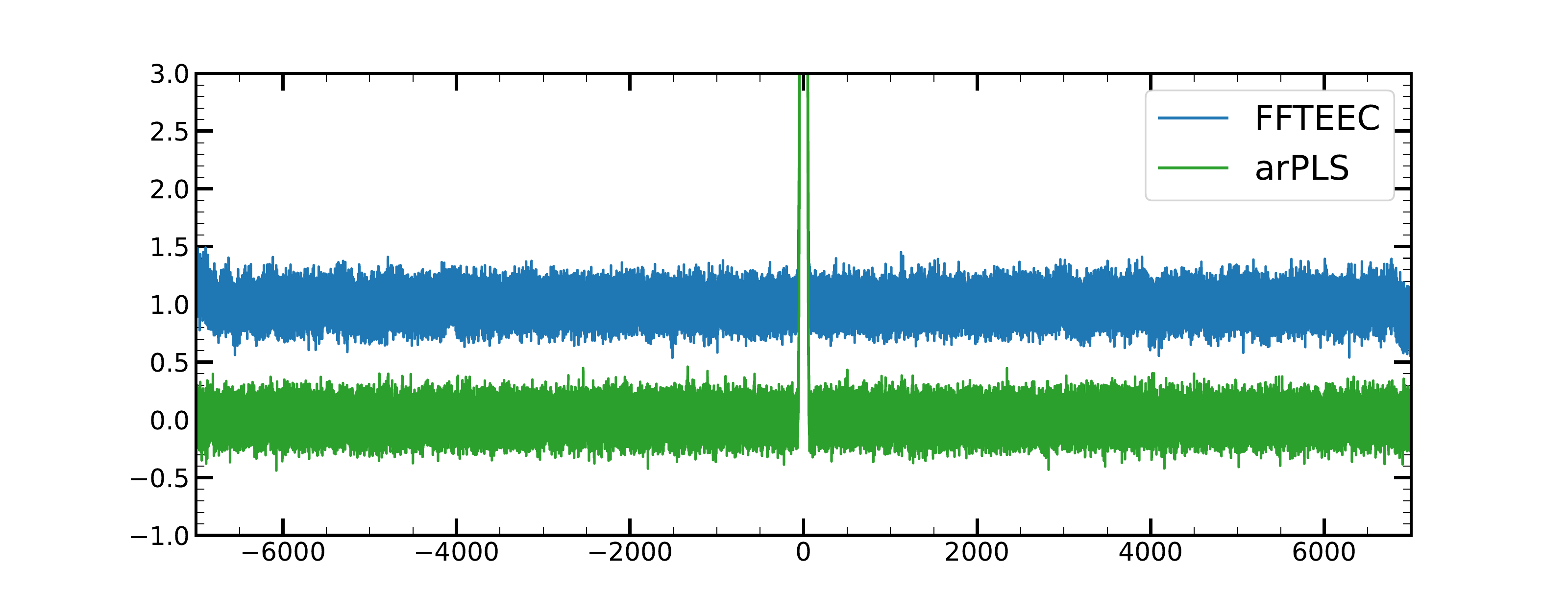}
    \includegraphics[width = 7cm]{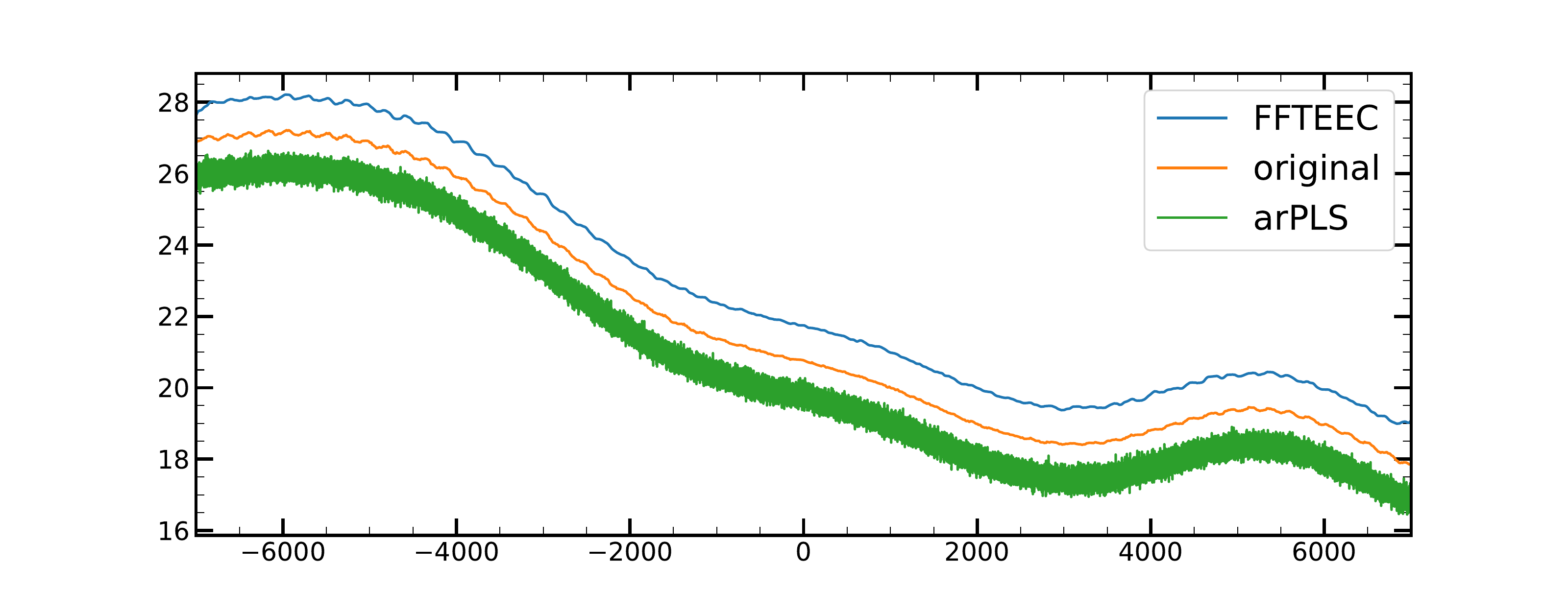}
    \caption{{\it Left}: Results processed by different methods. The blue line shows the FFTEEC result, where the zero point is 1, and the green line is the arPLS result. The flat baselines indicate that both methods can be used to correct inclined baselines. {\it Right}: Simulated standing waves and residuals from different methods. The blue line is the FFTEEC result and the green line is the arPLS result. The orange line is the simulated standing wave. The spectra have been separated by 1 on the $y$-axis to aid presentation. The profiles of the residuals of arPLS and FFTEEC are consistent with the simulate baseline. Standing waves are obvious in the residuals of FFTEEC, but unclear in the residuals of arPLS since the random noise is very high.}
    \label{fig:sim_com}
\end{figure}

Figures~\ref{fig:sim_com} displays the results and residuals processed by FFTEEC and arPLS, respectively~\footnote{We used the C++ arPLS software package provided by Ganriel Kronberger to speed up the calculations. \url{https://github.com/heal-research/arPLS}}. Both methods can be used to correct the inclined baseline, since the baselines in the left panel are sufficiently flat. The baseline obtained by arPLS is better than that obtained by FFTEEC on the two ends of the spectrum. In the right panel, although the profile of the residuals of arPLS are comparable to the simulated baseline, it is hard to judge whether the standing waves have been removed since the noise level of the residuals is very high. In contrast, the residuals from FFTEEC are remarkably consistent with the simulated baseline, not only with the profile, but also the standing waves. 

\begin{figure}[htbp]
    \centering
    \includegraphics[width = 8cm]{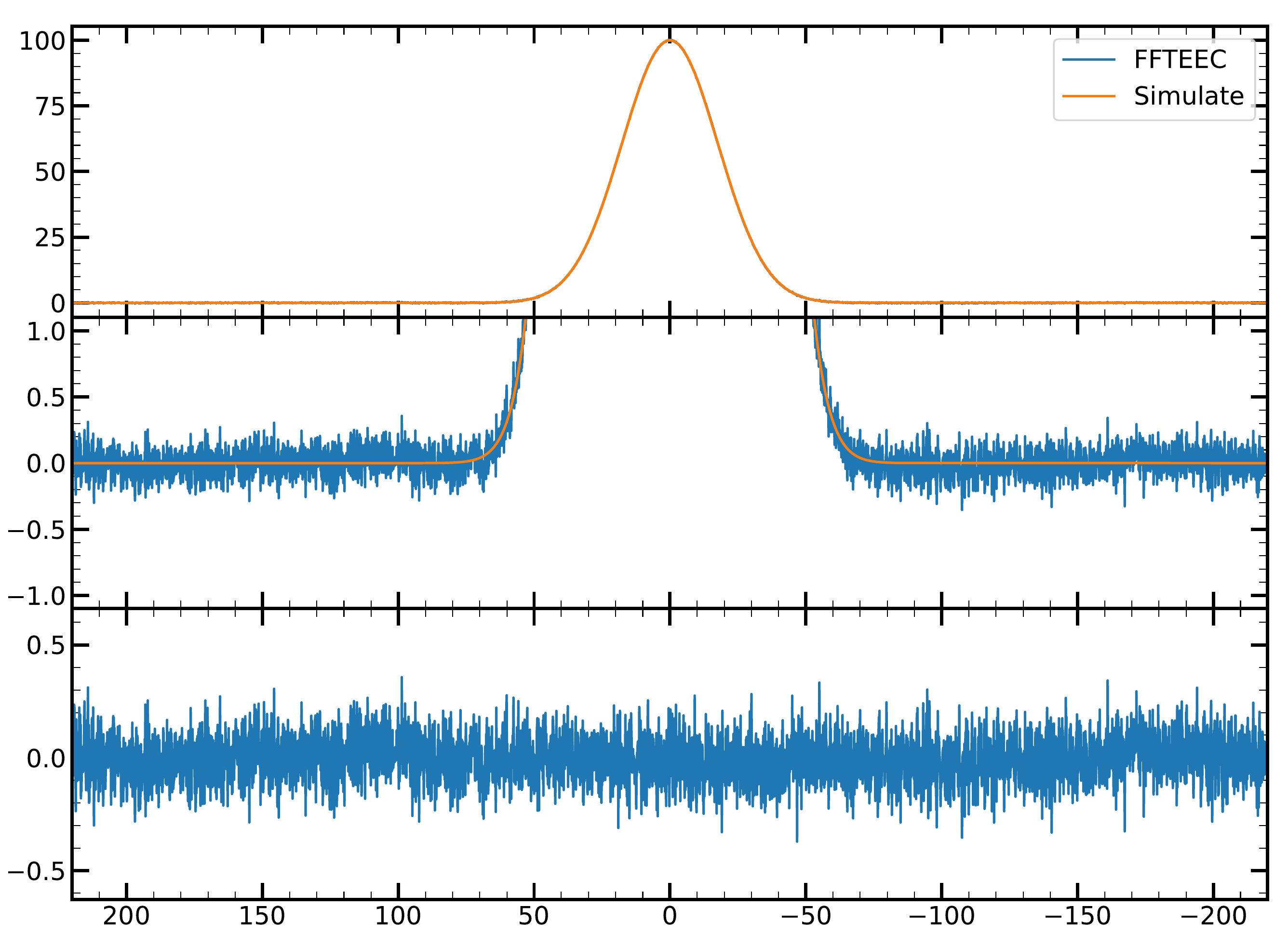}
    \caption{Results for simulated data. {\it Top}: The orange line is the simulate signal and the blue line is the spectra processed by FFTEEC. {\it Middle}: An enlarged figure of the top panel. {\it Bottom}: The residuals between the simulated signal and our result. The profile of simulated signal is similar to the results produced by FFTEEC in the top and bottom panels. The residuals are flat in the bottom panel, indicating that the standing waves have been removed from the baseline and we did not introduce any false signals or lose signals.}
    \label{fig:sim_FFT}
\end{figure}

Figure~\ref{fig:sim_FFT} displays a comparison between a simulated signal and spectra processed by FFTEEC, and it can be seen that all profiles are consistent (see details in the top panel). To show the differences more clearly, we have provided a partially enlarged view in the middle panel, and residuals are displayed in the bottom panel. The signals in the spectra processed by FFTEEC are consistent with the simulated signal, and the residuals between them are flat. In the simulated data, the width of the 3$\sigma$ signal is 60.2, while that obtained by the automatic signal extraction method is 84.8, which is wider than the signal. 

The root mean square error (RMSE) can be applied to illustrate the difference between two spectra, which can be expressed as:
 
\begin{equation}
    {\rm RMSE} = \sqrt{\frac{1}{N} \sum_{n = 1}^{N} [y(n) - {\widetilde{y}}(n)]^{2}},
\end{equation}
where $N$ is the length of the sequence, and $y$ and ${\widetilde{y}}$ are the signals to be compared. The RMSE between the simulated signals and the spectra processed by FFTEEC is 0.10, and that between the simulated signals and the simulate signals after random noise injection is also 0.10, indicating that there is only random noise in our result. As highlighted by the visual effect shown in Figure~\ref{fig:sim_FFT} and RMSE values of the simulated data, our method does not lose signals or introduce false signals.

The RMS can also be employed to judge the effectiveness of our method, which can be expressed as:

\begin{equation}
    {\rm RMS} = \sqrt{\sum_{n = 1}^{N}\frac{[y(n)]^{2}}{N}},
\end{equation}
where $N$ is the length of the sequence and $y$ is the baseline after removing the signal. We compared the RMS noise levels between the random noise of simulated data and our results, and the ratio between them is 1.12, indicating that our method can get a near-ideal RMS noise levels.

\subsection{Application to real data}
We applied a 10th order polynomial fitting method to correct the inclined baseline preliminarily. However, the baseline is roughly flat after polynomial correction, but still bumpy, and there are massive standing waves with different periods in it.

\begin{figure}[htbp]
    \centering
    \includegraphics[width = 7cm]{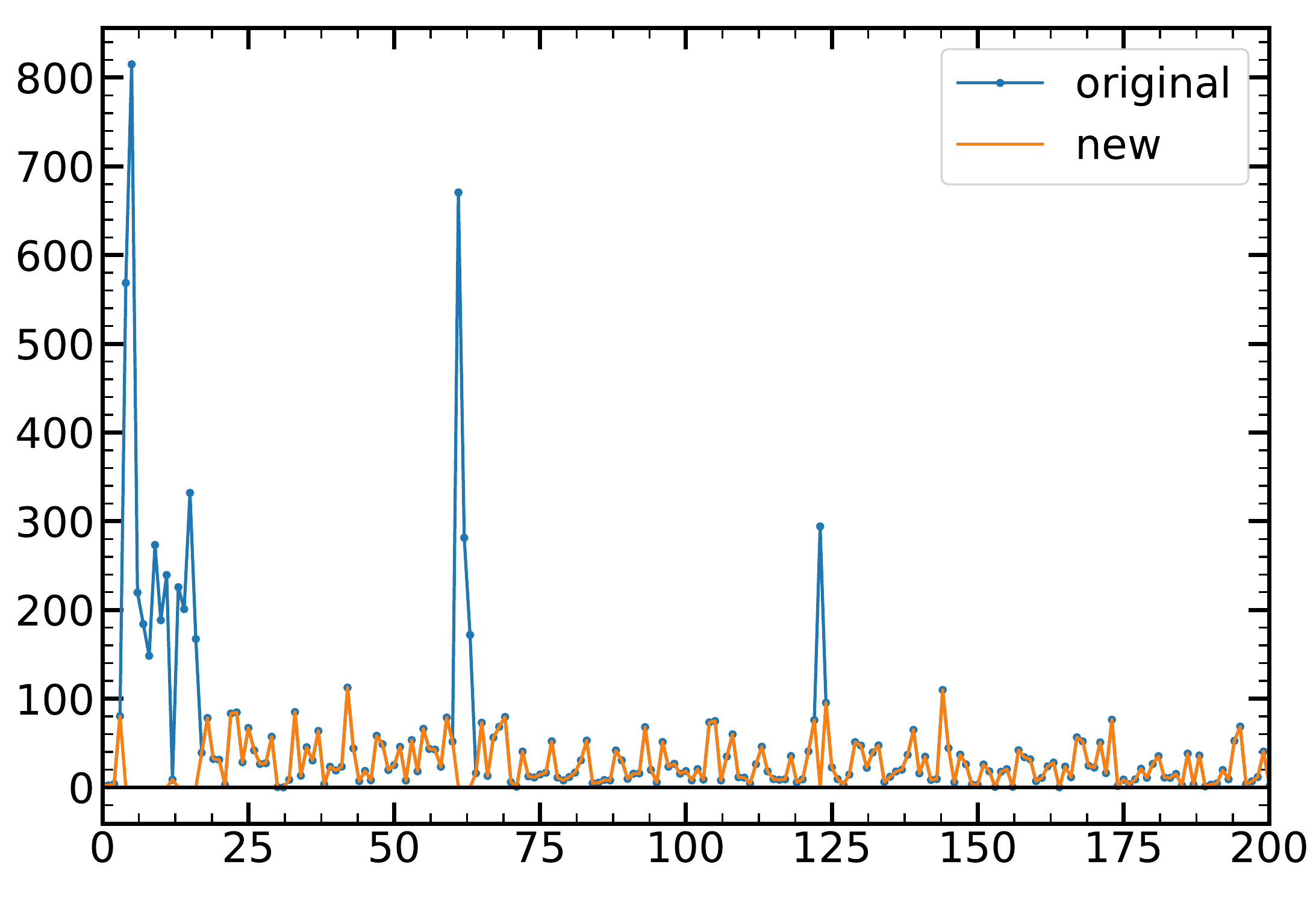}
    \caption{FAST baseline in the frequency domain. There are three obvious peaks in the blue line, corresponding to the frequencies of standing waves. The peaks have been removed in the orange line, indicating that our method can effective extract standing waves.}
    \label{fig:fourier}
\end{figure}

Figure~\ref{fig:fourier} presents the baseline of FAST in the frequency domain processed by FFT. The blue line is the original data, and there are three distinct standing waves with different frequencies. The orange line is the result where the standing waves have been removed. Here, we found that the frequencies of the standing waves of most FAST data are relatively consistent and contain three different frequencies, as shown in the figure after processing the data from different beams. However, some standing waves appear to have additional frequencies; e.g., the XX polarization of Beams M10, M13, and M15, the YY polarization of Beams M06, M07, and M17, etc. Taking the first 20 orders can remove almost all standing waves in the different data sets, whether the standing waves contain three frequencies or more. Thus the order of the FFT algorithm usually can be assigned to 20, indicating removal of the top 20 items with the largest amplitudes in the frequency domain. This number was obtained after experimenting using various FAST data sets.  

\begin{figure}[htbp]
    \centering
    \includegraphics[width = 7cm]{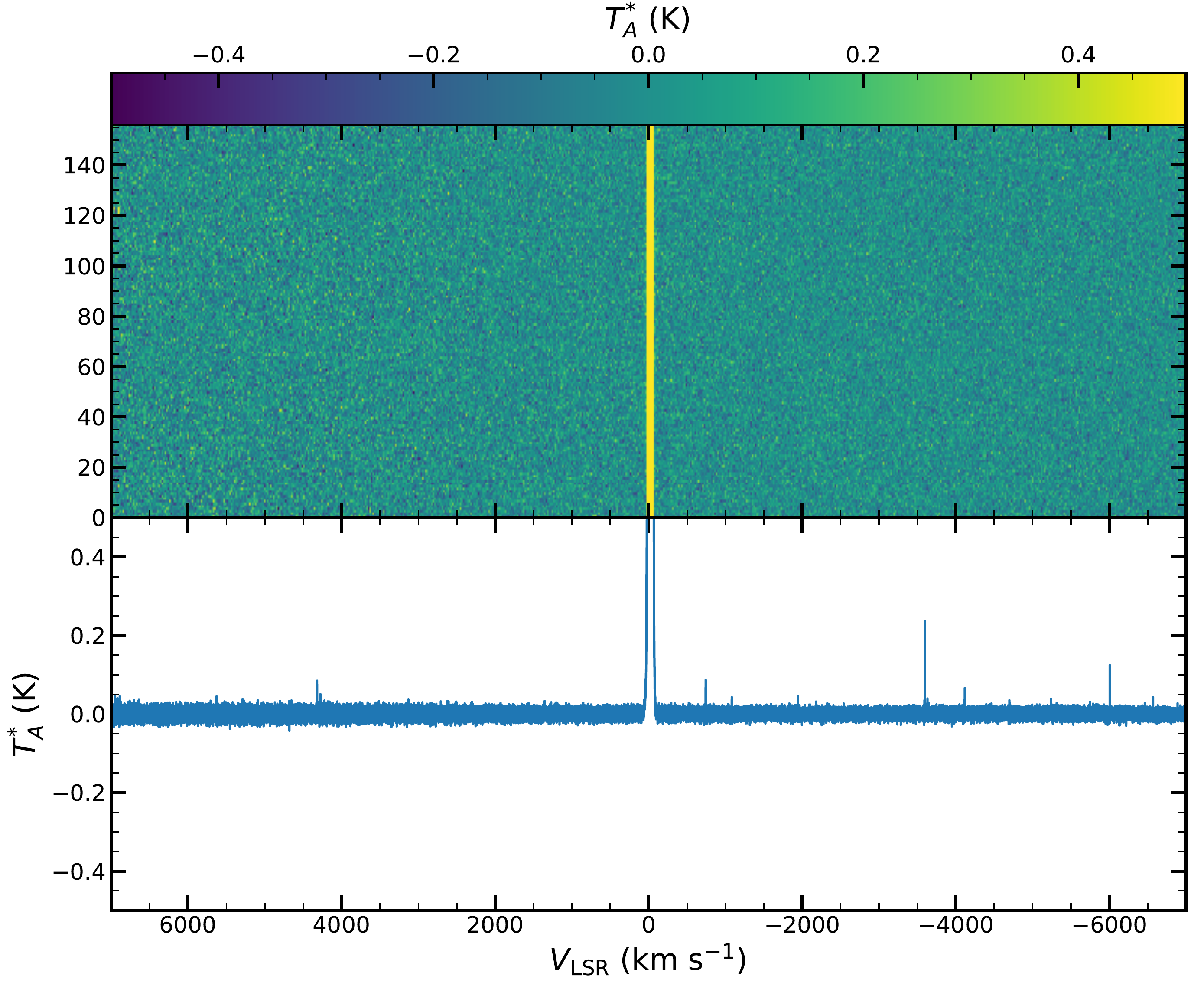}
    \includegraphics[width = 7cm]{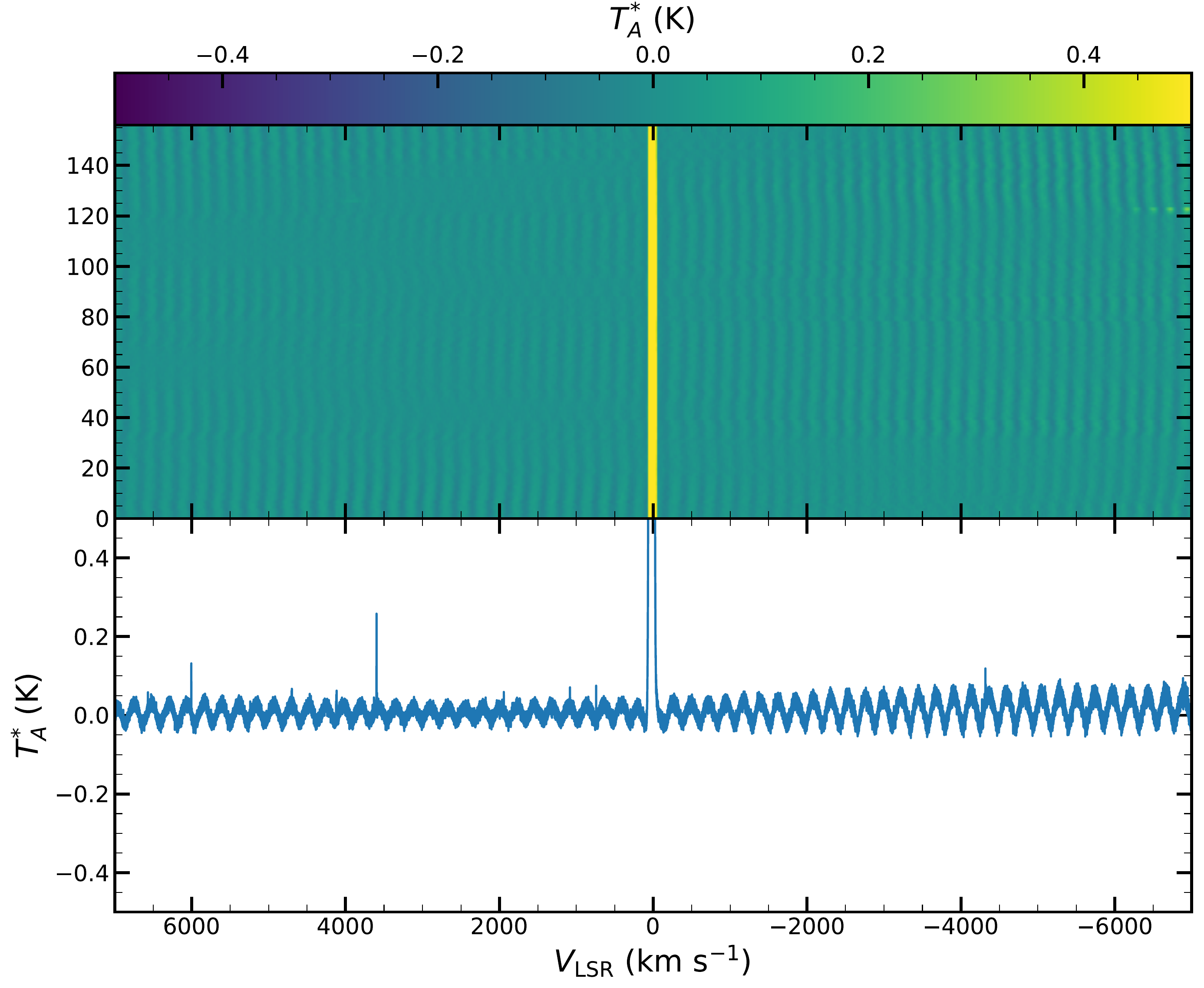}
    \caption{{\it Left}: Results of FFTEEC. FFTEEC can remove the standing waves in the original H~\textsc{i} spectra produced by FAST, since there is only random noise in the water-fall image, and the average spectrum is flat. {\it Right}: Results of arPLS. Standing waves are obvious in the figure, whether in the water-fall image or the average spectrum, indicating that arPLS is not good at removing standing waves.}
    \label{fig:res}
\end{figure}

Figure~\ref{fig:res} shows spectra corrected by FFTEEC and arPLS, respectively. In the left panel, the standing waves have been removed from the spectrum, since there is only random noise in the water-fall image. Meanwhile, FFTEEC is stable for the different spectra, since the average spectrum is so flat that it is hard to find any obvious standing waves. The result processed by arPLS is shown in the right panel; no matter in the water-fall image or average spectrum, standing waves are obvious, indicating that arPLS can just correct the inclined baseline and is not good at removing standing waves in the original H~\textsc{i} spectra produced by FAST. 

Table~\ref{tab:rms} lists the RMS noise of the average spectrum of five groups, i.e., the XX polarization of Beams M02 and M03 and the YY polarization of Beams M01, M06, and M13. The first row is for arPLS and the second row is for FFTEEC. The parameter $\lambda$, which is used to control the balance between fitness and smoothness, of arPLS is $10^{11}$ when real data were processed. Based on the RMS estimate method of the average spectrum provided by FAST~\citep[see Eq. 10 of ][]{Jiang+etal+2020}, the theoretical RMS of the average spectrum, integrated for 335 minutes, is 6 $\sim$ 8~mK. The RMS noise of arPLS is $\sim$20~mK, which is about three times greater than the theoretical RMS. As a comparison, the RMS noise of FFTEEC is $\sim$8~mK and approaches the theoretical RMS. 

In conclusion, although arPLS can be applied to correct the inclined baseline, it is hard to remove standing waves from the baseline. In contrast, FFTEEC can effectively remove standing waves and obtain a near-ideal RMS. There are some shortcomings for FFTEEC, e.g., it can only be used to correct the baseline when knowing the position of the signal, but it cannot automatically extract the signal as arPLS can. Additionally, sometimes the parameters of FFTEEC affect the results; e.g., the length of the smoothing box when automatically extracting a signal.

\begin{table}[htbp]
	\centering
	\begin{threeparttable}
	\caption{RMS noise of five groups of data}
	\label{tab:rms}
	\begin{tabular}{c|ccccc}
		\hline \hline
		Method & 1 & 2 & 3 & 4 & 5\\ \hline
		arPLS & 25~mK & 18~mK & 20~mK & 24~mK & 23~mK \\
		FFTEEC & 8~mK &9~mK & 9~mK & 8~mK & 9~mK \\
		\hline
	\end{tabular}
	\begin{tablenotes}
	\item {\bf Note.} The RMS noise of FFTEEC is $\sim$8~mK, and that of arPLS is $\sim$20~mK, which is significant higher than theoretical RMS if the standing waves have not been removed.
	\end{tablenotes}
	\end{threeparttable}
\end{table}

\normalem
\begin{acknowledgements}
This work was funded by NSFC Grant Nos. 11933011 and 11873019, the Natural Science Foundation of Jiangsu Province (Grants No. BK20210999) and the Key Laboratory for Radio Astronomy. This work used data from the Five-hundred-meter Aperture Spherical radio Telescope (FAST). FAST is a Chinese national mega-science facility, operated by the National Astronomical Observatories of Chinese Academy of Science (NAOC). L.Y.J. thanks to the support of the Jiangsu Province Entrepreneurship and Innovation Program.

\end{acknowledgements}

\bibliographystyle{raa}

\end{document}